\documentclass[11pt,epsfig,epsf]{article}
\usepackage{cite}
\newcommand{\ba}{\begin{array}}
\newcommand{\ea}{\end{array}}
\newcommand{\bd}{\begin{displaymath}}
\newcommand{\ed}{\end{displaymath}}
\newcommand{\be}{\begin{equation}} 
\newcommand{\ee}{\end{equation}}  
\newcommand{\bea}{\begin{eqnarray}}
\newcommand{\eea}{\end{eqnarray}}

\usepackage{times} 
\usepackage{setspace}

\usepackage{graphicx}
\def\q2 {q^2}

\begin{document}

\parindent 00pt

\begin{centering}

{\Large{\bf{Response of perturbed spherium to external fields }}}\\[9mm]
Poonam Silotia $^{1,}$\footnote{Corresponding author: psilotia21@gmail.com}
Bhawna Vidhani $^{2,}$\footnote{bhawna.vidhani@gmail.com} 
Vinod Prasad $^{3,}$\footnote{vprasad@ss.du.ac.in}
 


\parindent 00pt
{\em  $^{1}${Department of Physics and Astrophysics, University of Delhi, Delhi-110007, India}}\\
{\em  $^2${Department of Physics, Hansraj College, University of Delhi, Delhi-110007, India}}\\ 
{\em  $^3${Department of Physics, Swami Shraddhanand College, University of Delhi, Delhi-110036, India}}\\ 

\end{centering}

\vskip 17pt
  
{\bf {Abstract}} 
\doublespacing
We study the spectra and response of Hooke's law spherium (two electrons moving on a 
surface of a sphere and interacting via harmonic potential to external static and laser fields. The spectrum of the Hooke's law spherium is analysed
in the light of varying coupling strength. In addition, the $<cos^n\theta>$, $n=1,2$ for different
contributions of 'static' and 'laser' fields is studied and analysed. 


\vskip 1 true cm


\section {Introduction}
Recently much attention is being paid to the quasi-solvable systems such as Hooke's atom, two 
electrons on a sphere (spherium), ballium, jellium etc. due to their similarities to more complex
real quantum systems. Such systems are now being used as a prototype in order to understand many 
physical properties of quantum systems. Loos and McGill have devoted much work to the study of 
spectra of various two electron systems such as: two electron systems on hypersphere, sphere, 
two electron quantum rings, electrons on concentric spheres, etc.\cite{Loos1}-\cite{Loos8}.  As shown
by Loos et. al. \cite{Loos1}, two models of two electron systems namely  
harmonium or hookium, where electrons are bound by harmonic potential and spherium in which electrons 
are bound to move on the surface of the sphere, are quite useful to test various approximations \cite{Loos1} 
and references therein.) 

In this work, we focus on the spectra of Hooke's law spherium which consists of two interacting electrons.
The interaction between those two electrons is Coulombic and the electrons are forced to remain on the surface 
of the sphere \cite{Loos2}. In addition, as shown in ref. \cite{Loos1}, the electrons are interacting via a force constant.
In addition, we study the response of such spherium to external static and electromagnetic fields.
This kind of a system is not only valuable for the reasons mentioned above, but also a prototype for 
confined and/or hindered rigid rotor \cite{tyagi9}. The interaction term of the Hamiltonian in case of spherium, 
can be taken as a hindering potential in case of hindered rotor. Hence, we study the spectrum of the system, 
for various coupling strengths and obtain energy eigenvalues and various coupling matrix elements.

The response of spherium to external electric fields, is of importance in a number of research areas of 
atomic, molecular and chemical physics. This can be used to understand the stereodynamics of collisions 
involving molecules and manipulating molecules and atoms using external fields \cite{schmidt_JCP10}-\cite{schmidt_PRA12}.
The paper is organised as follows: in next section, 
we present necessary theoretical concepts and the computational method used to solve the problem, followed 
by results and discussion. Finally conclusion of the study is presented. 

\section{Theoretical Methods}

Let us consider two particles moving on the surface of a sphere of radius R. The coordinates of the particle 
on the surface of a sphere are given by $\Omega\equiv(\theta,\phi)$. The time independent Schr\"odinger equation
for two particles, on the surface of a sphere, is given by (in a.u.):

 \begin{equation}
 \left[\frac{1}{R^2}\nabla_1^2  + \frac{1}{R^2}\nabla_2^2 + 
 V(\Omega_1,\Omega_2)\right]\psi(\Omega_1,\Omega_2)= E\psi(\Omega_1,\Omega_2)
 \end{equation}
 where
 \begin{equation}
\nabla_i^2= - \left[\frac{1}{sin\theta_i} \left(\frac{\partial }
{\partial \theta_i} \sin\theta_i  \frac{\partial }{\partial \theta_i} \right)     
+  \frac{1}{sin^2\theta_i}    \frac{\partial^2 }{\partial \phi_i^2}  \right]
 \end{equation}

 and $V(\Omega_1,\Omega_2)$ is the interaction term.

 As reported by Aghekyan et al. [13], various models of interaction potential 
 such as Coulombic, Gaussian and many other forms of potentials, have been considered in literature for electrons
 moving on surface of sphere. However, here we have taken the interaction term  $V(\Omega_1,\Omega_2)$ as harmonic 
 and defined as $\alpha U$, 
  
 where 
 
 \be
 U\equiv R \sqrt{2-2 \cos \beta} 
 \ee
 
 and 
 \be
 \cos \beta= \frac{\vec r_1. \vec r_2}{R^2} = \cos\theta_1 \cos\theta_2 +\sin\theta_1 \sin\theta_2 \cos(\phi_1 -\phi_2)
 \ee
 and $\alpha$ is a constant.
 
 With little calculations it can be shown that the above equation can be written as \cite{Loos1}:
 \begin{equation}
 -\frac{1}{R^2}\left[ \frac{d^2}{d\theta^2} + \cot \theta \frac{d}{d \theta}\right]\psi(\theta) 
 + \alpha^2 R^2 \left( 1-\cos \theta  \right)\psi(\theta)= E'\psi(\theta)
 \end{equation}

 \begin{equation}
 \left[-B_e\left[\frac{d}{d\theta^2} + \cot \theta \frac{d}{d \theta}\right] 
 + \alpha^2 R^2 \left( 1-\cos \theta  \right)\right]\psi(\theta)= E'\psi(\theta)
 \end{equation}
 
 where $B_e$ is taken as constant = $\frac{1}{R^2}$
 
 \begin{equation}
 \left[-\left(\frac{d}{d\theta^2} + \cot \theta \frac{d}{d \theta}\right) + 
 \frac{\Omega'^2}{B_e}  \left( 1-\cos \theta  \right)\right]\psi(\theta)= \frac{E'}{B_e}
 \psi(\theta)
 \end{equation}
 
 where $\Omega'^2=\alpha^2 R^2$.
 
 \begin{equation}
 \left[-\left(\frac{d}{d\theta^2} + \cot \theta \frac{d}{d \theta}\right)
 + {\Omega^2}  \left( 1-\cos \theta  \right)\right]\psi(\theta)= E\psi(\theta)
 \end{equation}

 Solving above equation, we get energy eigenvalues of the system in units of $B_e$.
 We define $\Omega^2$ as coupling constant (in units of $B_e$). By varying $\Omega^2$, we  get 
 spectrum of the system  for given values of $\Omega^2$. In addition, this equation is similar
 to the equation of hindered rotation in some sense. To solve equation no. (8) we use finite difference method \cite{VP_BD14}-\cite{PS_CPB16}. 
 The response of such system to external field depends on the interaction matrix elements $<\psi|cos \theta|\psi'>$,
 hence we also evaluated, values of  $<\psi_j|cos \theta|\psi_{j'}>$ and $<\psi_j|cos^2 \theta|\psi_{j'}>$ for various values
 of coupling constant $\Omega^2$.
 
 When such a spherium is placed in external static and laser fields; it may be noted that for large values of $R$ (exceeding few hundred a.u.), the energy difference between successive states is in the range of $meV$ (as shown in Table-I), hence the total Hamiltonian of the system can be written as:
\be
H=H_0 + H^s_{int} + H^L_{int}
\ee

As mentioned earlier, all terms are represented in terms of $B_e$. $H^s_{int}$ is the interaction of 
static electric field with the spherium, while 
$H^L_{int}$ is the interaction of the laser field.

\be
H^s_{int}= -R E_s' cos \theta 
\ee
$E_s'$ is the static electric field strength, $E_s=R E_s'$. 

and

\be
H^L_{int}= -R E_L' cos \theta 
\ee
$E_L'$ is the strength of laser field, $E_L=R E_L'$. 


Hence time dependent Schr\"odinger equation for the spherium in static electric and electromagnetic field is given by (in a.u.):

\be
i \frac{\partial}{\partial t}\Phi= H\Phi=(H_0+H_s+H_L)\Phi
\ee

\be
i \frac{\partial}{\partial t}\Phi= (H_0-R E_s' \cos \theta - R E_L' \cos \theta \cos \omega t)\Phi
\ee

Since the $H=(H_0-R E_s' \cos \theta - R E_L' \cos \theta \cos \omega t)$ is time dependent with period $T=2 \pi/\omega$ where $\omega$ is the angular frequency of the laser field. For the case, where $\omega$ is near to resonance frequency and the intensity is not too high, the solution of the time dependent Schr\"odinger equation can be written, using rotating wave approximation (RWA) \cite{pramana17}

\be
\Phi_j=\sum\limits_{j=1}^n a_j \psi_j e^{-(i \lambda_j t)} e^{(-i j\omega t)}
\ee

where $n$ is the total number of states taken into account, $\psi_j's$ are the solutions of equation (8). Substituting equation (14) into equation (13) and using RWA we get a set of eigenvalue equation \cite{singhal18}, \cite{kriti19}.

The resulting equation can be written in matrix form. Now the resulting system is time independent. The quasi-energies $(\lambda_n)$ and the corresponding eigenvectors $(a_j)$ are evaluated by diagonalising the resulting equations. The transition probability for a transition from initial state $j$ to final state $j'$ are found by 

\begin{centering}
\be
P_{j \rightarrow j'} = |a_{j},a_{j'}|^2
\ee
\end{centering}


It is worth mentioning that as the quasi energy matrix is time independent and the frequency of the laser field taken in the study is $\omega=E_2-E_1$ which is very close to zero. Hence the dressed states are roughly time independent. So, we calculated the orientation and alignment in different dressed states as $<\Phi_j|cos \theta |\Phi_j>$ and $<\Phi_j|cos^2 \theta |\Phi_j>$.

\section{Results and Discussion}

Here, we have studied the response of a spherium to external static and laser fields. The radius of the spherium is taken to be 200 a.u. (i.e. $B_e= 2.5\times 10^{-5} a.u. $). The spectrum of such a spherium resembles to that of a hindered rotor. We solve the time independent Schr\"odinger equation using Finite Difference method. In addition, the coupling matrix elements $<\psi_j|cos^n \theta|\psi_{j'}>$, (where $n$= 1,2) are also evaluated. 

We present the energy eigenvalues (in units of $B_e$ in Table-I). As can be seen from the table for $\Omega^2=0$ (i.e. coupling constant zero) the energy eigenvalues of the spherium are exactly equal to that of the free rotor. With increase in $\Omega^2$, the energy eigenvalues change considerably for low lying states, however for higher states the variation in energy is small. This low variation is expected as perturbation of the energy levels will depend on the strength $\Omega^2$. 

The same can be said about the coupling matrix elements. For example, in Table-II the values for $<\psi_j|cos \theta|\psi_j'>$ are given for nearest neighbouring states. We can see that the coupling elements $<\psi_j|cos \theta|\psi_{j'}>$=0 for $\Omega^2$=0, as expected, however with increase in $\Omega^2$, these values show sharp increase. The variation is significant for low lying states. The trend is also reflected in $<\psi_j|cos^2 \theta|\psi_{j'}>$ as shown in Table-III. These matrix elements are required to study the response of spherium to external fields. 

The variation of energy eigenvalues (in units of $B_e$) of the lowest eight states
of a spherium with the static electric field (in units of $B_e$ )  is shown in Fig.1.
The coupling constant is varying along the column having three values 0, 1 and 8, and the laser field strength in units of $B_e$ varies along row with values of 0, 20 and 120. In the absence of laser field (fig. 1a, 1d and 1g) the energy levels show a shift with increase in the electric field i.e., showing a Stark shift. Compared to the Stark shift, the shift in energy due to increase in coupling constant is small as shown in Table-I. However, in the presence of laser field, the dressed states show many avoided crossing regions and all of the dressed states show significant red shift.

In Fig.2 is plotted the variation of probability of the lowest eight states of a
spherium with the static electric field (in units of $B_e$). The coupling constant is varying along the column having three values 0, 1 and 8, and the laser field strength in units of $B_e$ varies along row with values of 0, 20 and 120. When $E_L$=0, i.e., the first column, there is a change in the coupling constants and hence the change in the probability. The most affected coupling constant is $<0|cos \theta|1>$ resulting in a large change in the probability of the first excited state. Same can be said about the probability of other excited states. With increase in the strength of the laser field, oscillations set in the value of probability which increases with further increase in the strength of laser field.

Fig. 3 shows the variation of orientation parameter, $<cos \theta>$,  of the lowest eight states of a spherium with
the static electric field (in units of $B_e$). Along the column the coupling constant is increased and along the row the laser field strength in units of $B_e$ increases. In the absence of the laser field, the orientation parameter for higher values of electric field strength attains some constant value for different states e.g. in all the three cases $<cos \theta>_0$ i.e. orientation parameter of the ground state approaches 1, 0.92 for  $<cos \theta>_1$ and so on, which implies that the system is highly oriented even for higher values of coupling constants $\Omega^2$. However, with increase in the strength of laser field, this pattern is broken and the system goes from orientation to anti-orientation for large values of $E_L$.

The  variation of alignment parameter, $<cos^2 \theta>$ of the lowest eight states of a spherium with the 
static electric field (in units of $B_e$) is shown in Fig. 4.  As shown in previous figures, the coupling constant is varying along the column and the laser field strength in units of $B_e$ is varying along row. In the absence of laser field the alignment parameter for the ground state, $<cos^2 \theta>_0$, increases as the value of  $\Omega^2$ is increased approaching $\sim$ 1.0 for large values of static electric field. Further, for large values of the laser field strength, the system is highly aligned. 
Alignment parameter for the first excited state with small values of $\Omega^2$ first decreases and the increases with increase in $E_s$. For higher values of $\Omega^2$, it shows lower values at $E_s=0$ and then increases. However, its value remains approximately same for higher values of static electric field with increase in the coupling constant. For higher excited states, the alignment parameter shows variations with $\Omega^2$. For the laser field strength $E_L/B_e$=20, the first and second excited state is affected most, with increase in the coupling constant. For $E_L/B_e$=120, the lowest three states attain almost constant values in strong electric field region. Higher excited states show oscillatory behavour.

%
{\bf{Conclusion}}

In the present study, we have calculated the energies and the coupling matrix elements of first and second order, of a spherium. The energies and the coupling matrix elements show strong dependence on the strength of the interaction term ($\Omega^2$), due to which the system shows interesting behaviour to external fields, shown in terms of variation of the dressed states, transition probability, orientation parameter and alignment.

{\bf{Acknowledgement}}

One of us (PS) is grateful to University of Delhi for providing the funds under the 'Scheme to strengthen Reasearch and Development'.

\bibliographystyle{unsrtnat}

\begin{thebibliography}{99}
\singlespacing
\bibitem{Loos1} P.-F. Loos, Phys. Rev. A {\bf{81}}, 032510 (2010).
\bibitem{Loos2} P.-F. Loos and  P.M.W. Gill, Chem. Phys. Lett. {\bf{500}}, 1 (2010).
\bibitem{Loos3} P.-F. Loos and  P.M.W. Gill, Phys. Rev. A {\bf{81}}, 052510 (2010).
\bibitem{Loos4} P.-F. Loos and  P.M.W. Gill, Phys. Rev. A {\bf{79}}, 062517 (2009).
\bibitem{Loos5} P.-F. Loos and  P.M.W. Gill, Phys. Rev. Lett. {\bf{105}},113001 (2010).
\bibitem{Loos6} P.-F. Loos and  P.M.W. Gill, Mol. Phys. {\bf{108}}, 2527 (2010).
\bibitem{Loos7} P.-F. Loos and  P.M.W. Gill, J. Chem. Phys. {\bf{132}}, 234111 (2010).
\bibitem{Loos8} P.-F. Loos and  P.M.W. Gill, J. Chem. Phys. {\bf{131}}, 241101 (2009).
\bibitem{tyagi9} B. Dahiya, A. Tyagi and V. Prasad,  Mol. Phys. {\bf{112}},  1651 (2014).
%
 \bibitem{schmidt_JCP10} B. Schmidt and B. Friedrich,  J. Chem. Phys. {\bf{140}}, 064317 (2014).
 \bibitem{sharma11} K. Sharma and B. Friedrich,  New J. Phys. {\bf{17}}, 045017 (2015).
 \bibitem{schmidt_PRA12} B. Schmidt and B. Friedrich, Phys. Rev. A  {\bf{91}}, 022111 (2015).
\bibitem{aghekyan13} N. G. Aghekyan, E. M. Kazaryan and  H. A. Sarkisyan, Few-Body Syst. {\bf{53}}, 505 (2012).
%

\bibitem{VP_BD14} V. Prasad and B. Dahiya, Physica Status Solidi {\bf{248}}, 1727 (2011).
 \bibitem{BD_JL15} B. Dahiya, V. Prasad and K Yamashita, J. Luminescence {\bf{136}}, 240 (2013).
\bibitem{PS_CPB16} P. Silotia, R. K. Meena and V. Prasad, Chin. Phys. B {\bf{24}}, 020303 (2015).
%
\bibitem{pramana17} N. Singhal, V. Prasad and M. Mohan, Pramana: J. of Phys. {\bf{62}}, 883 (2004).
\bibitem{singhal18} N. Singhal, V. Prasad and M. Mohan, European Physical Journal D {\bf{21}}, 293 (2002).
\bibitem{kriti19}  K. Batra, V. Prasad and M. Mohan, European Physical Journal D {\bf{20}}, 191 (2002).

\end{thebibliography}

\pagebreak

\begin{table}
\caption{Energies  of few lowest states (in units of $B_e$)  of spherium for different  values of   ${\Omega^2}$}
\centering 
\setlength{\tabcolsep}{4pt}

\begin{tabular}{ c c c c c c c c c} 
\hline 
\hline 
$Energy \hspace{1mm} states$ & ${\Omega^2=0}$  &${\Omega^2=0.1}$ &${\Omega^2=0.2}$  & ${\Omega^2=0.5}$ & ${\Omega^2=1}$ & ${\Omega^2=3}$ &  ${\Omega^2=5}$ & ${\Omega^2=8}$   \\          
          \hline 
$E_1$  & 0.00000 &0.09832 & 0.19333 & 	0.45892 & 0.84231	 &  1.90732 &  2.63438 & 3.480090 \\
$E_2$    & 2.00000   & 	 2.10097   &  2.203942 &	2.52426	&	3.09057	 & 5.47768 &		7.63857	& 10.26391  \\
$E_3$    & 6.00000   &	 6.10020   &  6.2008846 &	6.50579	&	7.02369	 & 9.226819	 &	11.64412&  15.44325 \\
$E_4$    & 12.00000  & 12.10006   &	 12.20034 &	12.50254 &	13.01064	 & 15.09957 &	17.28377&	 20.76488 \\
$E_5$    & 20.00000  &	 20.10000  & 20.20013& 	20.50131	&	21.00588	 & 23.05675	 &	25.16048& 28.41905  \\
$E_6$    & 30.00000  &	 30.09996  & 30.2000 &	30.50069	&	31.00352	 & 	33.03625  &	35.10340&	38.26954  \\	
$E_7$    & 42.00000  &	 42.09994  &  42.19994 &	42.50031	&	43.00214 &  45.02463 & 47.07144	& 50.18754 \\	    	
$E_8$    & 56.00000  & 56.09992   & 56.19988 &   56.50005 	& 57.00124 &	 59.0173	 & 61.05150	& 64.13691  \\
$E_9$    & 72.00000  & 72.099902  & 72.19983 &	72.49986 & 73.00060	 & 75.01233 &	77.03811& 80.10317 \\																			
$E_{10}$ & 90.00000  & 90.09988   & 90.19979 &   90.49970 	& 91.00011 & 	93.00874 & 95.02858	&  98.07939 \\
$E_{11}$ & 110.00000 & 110.09986  &  110.19976 & 110.49957	&	110.99972 & 	 113.00603 &	115.02150	&118.06188 \\									
$E_{12}$ & 132.00000 & 132.09985  & 132.19972 & 132.49946 &	132.99940 & 135.00391	 &	137.01604& 140.04853 \\
$E_{13}$ & 156.00000 & 156.09983  & 156.19969 & 156.49935	&	156.99912 & 159.00219 &		161.01169&  164.03804\\
$E_{14}$ & 182.00000 & 182.09982  & 182.19966 & 	182.49926&	182.99887 & 	 185.00075	 &	187.00815& 190.02959 \\
$E_{15}$ & 210.00000 &  210.09981 &  210.19963 &10.49917 &	210.99865	 & 212.99953 &	215.00519	&	218.02263  \\
$E_{16}$ & 240.00000 &  240.09979 &  240.19960 & 240.49909		&	240.99844  & 242.99846	 &	245.00266&  248.016799 \\   
$E_{17}$ & 272.00000 & 272.099784 &  272.19957 & 272.49901	&  272.99825 & 	274.99752 &		277.00048	& 280.01182	 \\
$E_{18}$ & 306.00000 &  306.09977 &  306.19954 & 306.49893	& 306.99807	 & 	308.99668	 &	310.99855	&	 314.00750 \\
$E_{19}$ & 342.00000 &  342.09975 &  342.19952 & 342.49886	& 342.9979 & 	344.99591 &		346.99684	&  350.003724 \\
$E_{20}$ &   379.00000                    & 380.09974 &  380.19949 &	380.49878	 & 380.99774 &	 382.99520 &384.99529	&  388.00036 \\

  \hline 
\end{tabular}
\label{evldr0} 
\end{table}
\pagebreak
\begin{table}

\caption{$<\psi_j\mid \cos \theta\mid \psi_{j'}>$ in spherium} 
\centering 
\setlength{\tabcolsep}{4pt}
\begin{tabular}{ c c c c c c c c c} 
\hline 
 
$Transitions$ & ${\Omega^2=0}$  &${\Omega^2=0.1}$ &${\Omega^2=0.2}$  & ${\Omega^2=0.5}$ & ${\Omega^2=1}$ & ${\Omega^2=3}$ &  ${\Omega^2=5}$ & ${\Omega^2=8}$   \\
          
\hline 

0-0 & 0.00000 &  0.03322  & 0.06620  & 0.16148 & 0.29873 & 0.57908 & 	0.67811	&  0.74682 \\								
0-1 & 0.57629 & 0.57560 &  0.57356 & 0.55997	 & 0.51963 &   0.35765 &  0.28334	 &  0.23040 \\
0-2 & 0.00000 & 0.00743 &  0.01479 & 0.03581	 & 0.06450 &  0.09489 &   0.07209 &  0.04237 \\
0-3 & 0.00000 & 0.16388 &  0.00009 & 0.00059 &  0.00215 &  0.00919 & 0.01151 &  0.00965 \\
0-4 & 0.00000 & 0.00000 & 0.00000 & 0.00000   & 0.00003   & 0.00039 & 0.00082	 &  0.00110 \\
0-5 & 0.00001 & 0.00001 & 0.00001 & 0.00001 & 0.00001 & 0.00000 & 0.00002 &   0.00006 \\
0-6 & 0.00000  & 0.00000 &  0.00000 & 0.00000   & 0.00000 &  0.00001 &  0.00001 &  0.00001 \\
0-7 & 0.00001 & 0.00001 & 0.00001 & 0.00001 & 0.00001  &   0.00001 &  0.00001 &  0.00002 \\
1-1 & 0.00000 & 0.01992 & 0.03960 & 0.09488	 & 0.16483 & 0.16359 & 0.00861 & 0.22235 \\ 		
1-2 & 0.51553 &  0.51550 & 0.51542 & 0.51486 &  0.51361 & 0.51229 & 0.48569  & 0.41276	 \\ 
1-3 & 0.00000 &  0.00218 & 0.00436 &  0.01092 & 0.02200 &  0.07000 &  0.11278  & 0.13015 \\
1-4 & 0.00002 &  0.00001 &  0.00000 &  0.00007 &  0.00034 & 0.00348 &  0.00954 & 0.01798 \\
1-5 & 0.00000 &	  0.00000 &  0.00000 &   0.00000 &  0.00000 &  0.00009 & 0.00042	 & 0.00127 \\ 
1-6 & 0.00002 &  0.00002 & 0.00002 &  0.00000 &  0.00002 &  0.00002 &   0.00000 & 0.00004 \\
1-7 & 0.00000 &  0.00000 & 0.00000 &  0.00000 &  0.00000 &  0.00000 & 0.00000 & 0.00001 \\
2-2 &  0.00000 & 0.00475 & 0.00952 & 0.02389	 & 0.04842	 & 0.15750 & .25108 & 0.25163 \\					
2-3 & 0.50633 &  0.50632 & 0.50631  & 0.50624	 &  0.50595 &  0.50251 &    0.49706 & 0.49344  \\
2-4 & 0.00000 &	  0.00106 & 0.00213 & 0.00532 &  0.01065 &  0.03228 &  0.05523 & 0.09454 \\
2-5 & 0.00002 &  0.00002 & 0.00002 & 0.00000 & 0.00008	 &  0.00091 &   0.00269 & 0.00765 \\
2-6 & 0.00000 &   0.00000 & 0.00000 &  0.00000 & 0.00000	 &  0.00002 &   0.00008 & 0.00036 \\
2-7 & 0.000003 &  0.00002 & 0.00003 & 0.00003 & 0.00003 &  0.00003 & 0.00002 &  0.00001 \\
3-3 & 0.00000 & 0.00003 & 0.00444  & 0.01110 & 0.02224 & 0.06794 & 0.11796	 &	0.20532 \\				 	
3-4 & 0.50327 & 0.50328  & 0.50328 &  0.50325 &  0.50317 &  0.50231 &  0.50041 &  0.49499 \\
3-5 & 0.00000 & 0.00063  &  0.00126 & 	0.00316 & 0.00633  &  0.01904 &  0.03193 &   0.05194 \\
3-6 & 0.00003 &	  0.00003  & 0.00003 & 0.00002 & 0.00000 &  0.00033 &  0.00099 &  0.00266 \\
3-7  & 0.00000 &  0.00000  & 0.00000 & 0.00000 & 0.00000 &  0.00000 & 	0.00002 &  0.00009 \\
4-4 & 0.00000 &	  0.00130  & 0.00259  & 0.00649 & 0.01298	 & 0.03912 & 0.06583  & 0.10816 \\			
4-5 & 0.50192  &  0.50193  &  0.50193 & 0.50192 & 0.50189 & 0.50156 &	  0.50089 &  0.49914 \\
4-6 & 0.00000 &  0.00042 & 	 0.00084 &  0.00210 & 0.00420 &  0.01261 &  0.02108 &  0.03394 \\
4-7 &  0.00003 &  0.00003 & 0.00004 & 0.00003 &  0.00002 & 0.00013  & 0.00044 & 0.00120 \\
5-5 & 0.00000 &	  0.00085 & 0.00171 & 0.00427	 & 0.00854 & 0.02567 & 0.04293	 &  0.06933	\\	
5-6 & 0.50123  &  0.50124 &  0.50124 &  0.50123 &  0.50122 & 0.50106 & 0.50076	 &  0.49998 \\
5-7 & 0.00000 & 0.00030 & 0.00060 &  0.00150 & 0.00299 & 0.00898 &0.01499	 & 0.02406 \\
6-6 & 0.00000 &  0.00060 & 0.00121 & 0.00303	 & 0.00606 & 0.01818 & 	0.03036	 & 0.04878 \\	

6-7 & 0.50085 &  0.50085 &  0.50085 & 0.50085 & 0.50084 & 0.50076 & 0.50060 & 0.50020 \\
7-7 & 0.00000 & 0.00045 & 0.00090 &  0.00226 & 0.00452 & 0.01357 & 0.02264  & 	0.03631 \\

  \hline 
\end{tabular}
\label{evldr0} 
\end{table}

\pagebreak

\begin{table}

\caption{$<\psi_j\mid \cos ^2\theta\mid \psi_{j'}>$ in spherium} 
\centering 
\setlength{\tabcolsep}{4pt}
\begin{tabular}{ c c c c c c c c c} 
\hline 
 
$Transitions$ & ${\Omega^2=0}$  &${\Omega^2=0.1}$ &${\Omega^2=0.2}$  & ${\Omega^2=0.5}$ & 
${\Omega^2=1}$ & ${\Omega^2=3}$ &  ${\Omega^2=5}$ & ${\Omega^2=8}$   \\
          
\hline 

0-0 &  0.33269 & 0.33306 & 0.33415 & 0.34152 & 0.36407 & 0.47327 & 0.54654 & 	0.61395 \\
0-1 &  0.00000 & 0.01150 &  0.022926 &  0.05585 & 0.10296 & 0.19824 & 0.23135	 & 0.24255 \\
0-2 & 0.29761 &  0.29747 & 0.29703 & 0.29405 &  0.28463 & 0.22830 & 0.17452 &  0.12119 \\
0-3 &  0.00000  & 0.00502 & 0.01000 & 0.024373 &  0.04475 & 0.07776 & 0.07479 & 0.05678	\\
0-4 &  0.00000 & 0.00001 & 0.00008 & 0.00053 & 0.00196 & 0.00916 &  0.01298 & 0.01369 \\
0-5 &  0.00000 & 0.00000 & 0.00000 & 0.00000 &  0.00003 & 0.00049 & 0.00110  & 0.00171\\
0-6 &  0.00001 & 0.00002 & 0.00001 & 0.00001 & 0.00001 & 0.00000 & 0.00004 & 0.00011 \\
0-7 & 0.00000 &  0.00000 & 0.00000 & 0.00000 & 0.00006 & 0.00001 & 0.00001 &  0.00001 \\
1-1 & 0.59895 & 0.59853 & 0.59727 &  0.58881	& 0.56246 & 0.42277 &		0.32964 &	0.29070\\
1-2 &  0.00000 & 0.00735 & 0.01465 & 0.03562 & 0.06498 & 0.09543 &  0.04081 & 0.06372 \\ 
1-3 &  0.26143 & 0.26138 &  0.26124 & 0.26027 & 0.25745 & 0.24666 & 0.23754 & 0.21749				  \\
1-4 &  0.00000 & 0.00164 &  0.00329 & 0.00823 &  0.01652 & 0.05125 & 0.08328 &  0.01916 \\
1-5 &  0.00002 & 0.00001 & 0.00000 & 0.00007 &  0.00035 &  0.00354 & 0.00970 & 0.10656 \\ 
1-6 &  0.00000 &  0.00000 & 0.00000 & 0.00000 &  0.00000 & 0.00012 & 0.00055 & 0.00174\\
1-7 &  0.00002 & 0.00002 &  0.00002 & 0.00002 &  0.00002 & 0.00001 & 0.00000 & 0.00008	 \\
2-2 & 0.52297&  0.52302 & 0.52315 &  0.52407	& 0.52724 &	0.55072 & 0.55519 & 0.48879 \\
2-3 &   0.00000 & 0.00187 & 0.00375 & 0.00940 & 0.01897 & 0.06040 & 0.10024 &  0.12435	\\
2-4 &  0.25517 & 0.25517 & 0.25515 &  0.25501 & 0.25451 & 0.24870 & 0.23762 & 0.22186	 \\
2-5 &  0.00000 & 0.00085 & 0.00171 & 0.00427 & 0.00855 & 0.02567 &  0.04304 & 0.07116\\
2-6 &  0.00003 &  0.00002 & 0.00002 & 0.00000 & 0.00009 & 0.00103 & 0.00299 & 0.00827 \\
2-7 & 0.00000 & 0.00000 & 0.00000 & 0.00000 & 0.00000 & 0.00002 & 0.00011 & 0.00051 \\
3-3 & 0.51038 & 0.51038 & 0.51040 & 0.51051& 0.51091 & 0.51553 & 0.52604 &	0.55124 \\							 	
3-4 &  0.00000 & 0.00091 & 0.00183 & 0.00457 & 0.00916 & 0.02779 &  0.04747 & 0.08024 \\
3-5 &  0.25291 & 0.25291 & 0.25290 & 0.25286 & 0.25271 & 0.25111 &  0.24771 & 0.23839 \\
3-6 &  0.00000 & 0.00052 & 0.00105 & 0.00264 & 0.00528 & 0.01587 & 0.02646 & 0.04236  \\
3-7  & 0.00003 & 0.00003 & 0.00003 & 0.00002 & 0.00001 & 0.00040 &  0.00119  &  0.00316 \\
	
4-4 & 0.50585 &  0.50585 &	0.50586 & 0.50589 & 0.50601 & 0.50725 & 0.50986 & 	0.51694	\\		
4-5 & 0.00000  & 0.00055 & 0.00110 & 0.00275 & 0.00551 &  0.01659 & 0.02781 & 0.04518 \\
4-6 &  0.25184  &0.25184 & 0.25184 &  0.25183 & 0.25182 & 0.25176 &  0.25111	 &  0.24980  \\
4-7 & 0.00000  &  0.00036 & 0.00072 &  0.00180 & 0.00360 & 0.01082 &  0.01804 & 0.02889 \\
5-5 & 0.50372 & 0.50372 & 	0.50372 & 0.50373 & 0.50378	& 0.50428 &
 0.50529 & 0.50783 \\	
5-6 &  0.00000 & 0.00037 & 0.00074 & 0.00185 & 0.25122 & 0.01115 & 0.01862 & 0.02995 \\
5-7 & 0.25125 &  0.25125 & 0.25125 & 0.25125 & 0.00371 & 0.25090 & 0.25027 & 0.24870 \\
6-6 & 0.50256 & 0.50256 &	0.50256 & 0.50256 & 0.50259 & 0.50283 &	0.50331	& 0.50451	\\	
6-7 & 00.00000  & 0.00026 & 0.00053 &  0.00133 & 0.00267 & 0.00804 & 0.01341  &  0.02151\\
7-7 &  0.50187 & 0.50187 & 0.50187 & 0.50188 &  0.50189 & 0.50202 & 0.50228 &  0.50293 \\

  \hline 
\end{tabular}
\label{evldr0} 
\end{table}
	
\pagebreak

\begin{figure}
\begin{center}
\scalebox{0.70}{\includegraphics{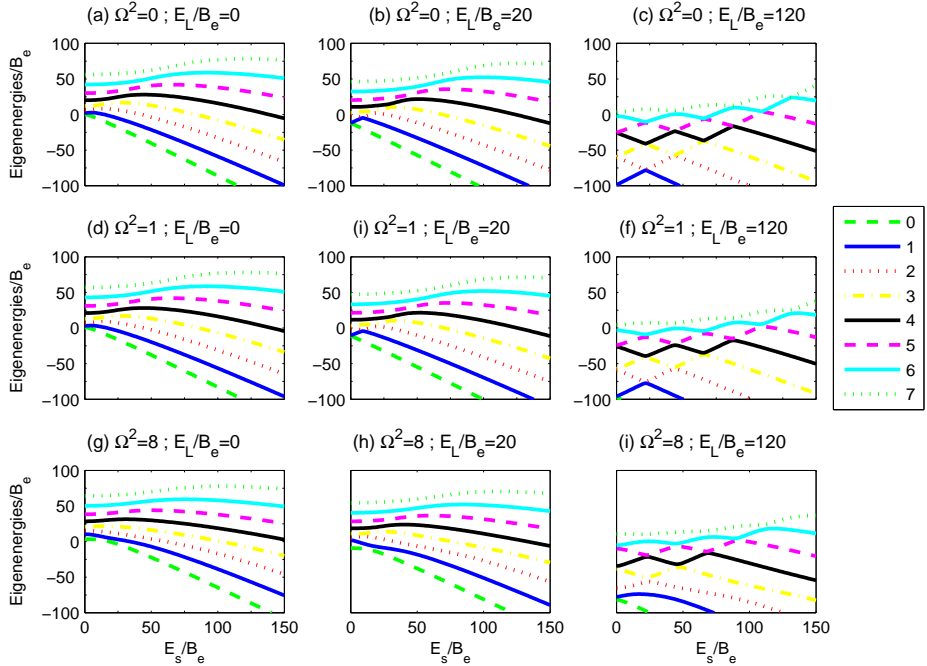}}
\caption{\em {Variation of energy eigenvalues (in units of $B_e$) of the lowest eight states of a spherium with the static electric field (in units of $B_e$).}}
\end{center}
\label{fig:fig1}
\end{figure}

\begin{figure}
\begin{center}
\scalebox{0.70}{\includegraphics{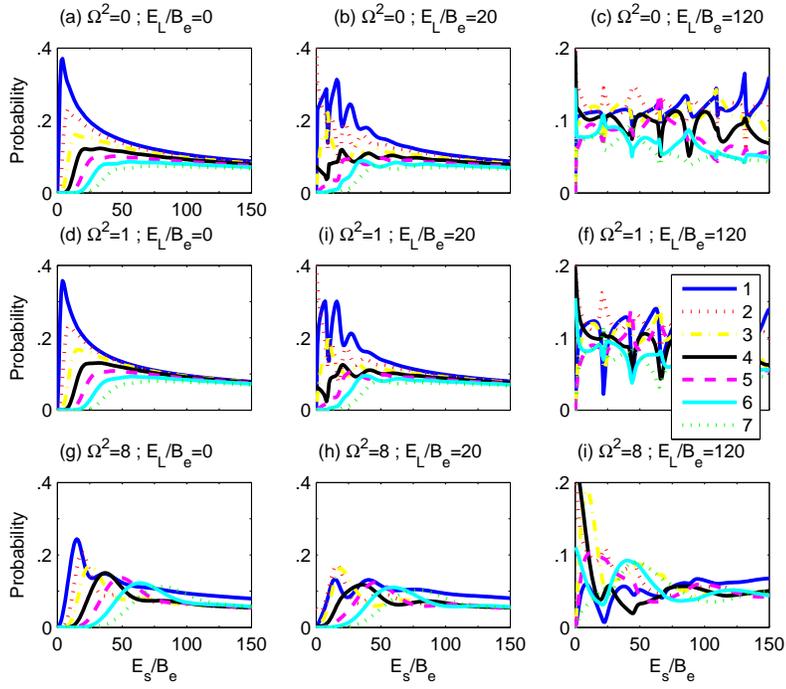}}
\caption{\em {Variation of probability of the lowest eight states of a spherium with the static electric field (in units of $B_e$).}}
\end{center}
\label{fig:fig2}
\end{figure}

\begin{figure}
\begin{center}
\scalebox{0.70}{\includegraphics{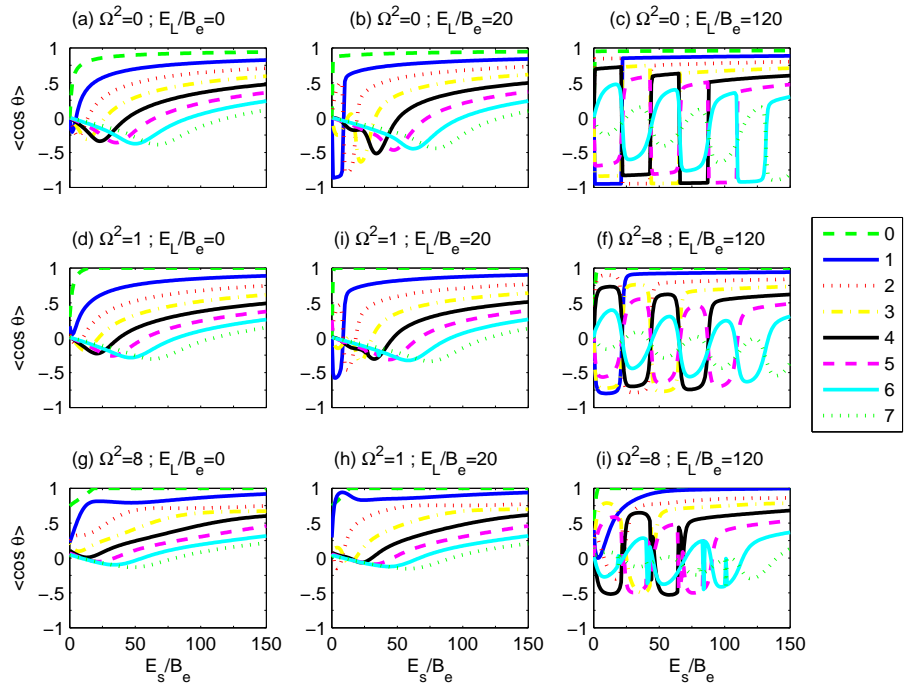}}
\caption{\em {Variation of orientation parameter, $<cos \theta>$, of the lowest eight states of a spherium with the static electric field (in units of $B_e$).}}
\end{center}
\label{fig:fig3}
\end{figure}

\begin{figure}
\begin{center}
\scalebox{0.70}{\includegraphics{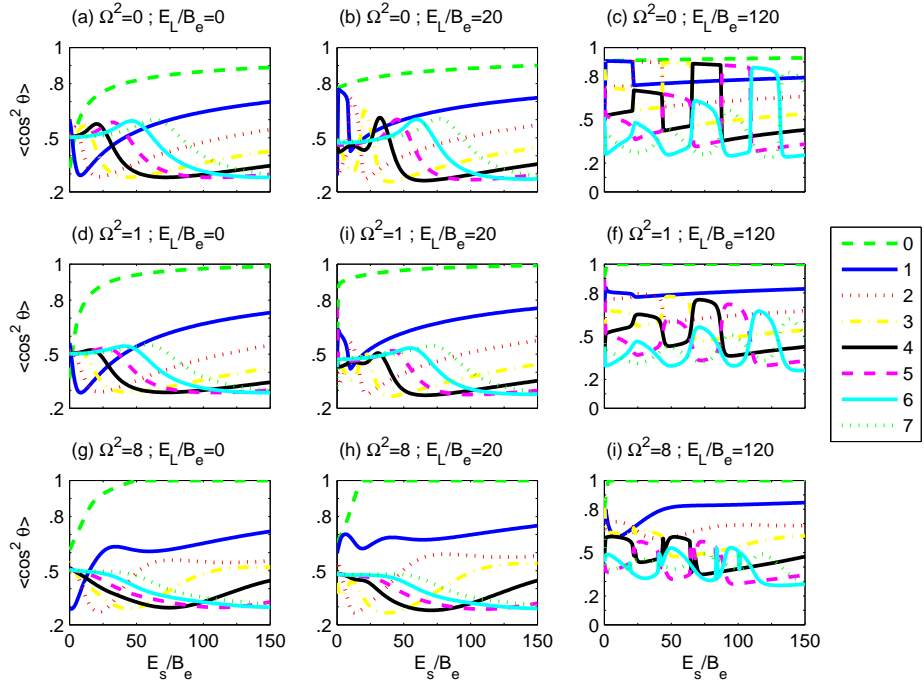}}
\caption{\em {Variation of alignment parameter, $<cos^2 \theta>$, of the lowest eight states of a spherium with the static electric field (in units of $B_e$).}}
\end{center}
\label{fig:fig4}
\end{figure}

\end{document}